\documentclass{PoS}

\usepackage{epsfig,macros_static,macros_ssm}
\usepackage{amssymb,fontenc,times,mathptmx,graphicx}

\title{$m_{\rm b}$ and $f_{\rm B_s}$ from a combination of HQET and QCD}

\ShortTitle{$m_{\rm b}$ and $f_{\rm B_s}$ from a combination of HQET and QCD}

\author{\speaker{Damiano Guazzini}, Rainer Sommer\\
        \\
        DESY,
        Platanenallee 6,
        D-15738 Zeuthen,
        Germany
        \\
        E-mail: \email{Damiano.Guazzini@desy.de, Rainer.Sommer@desy.de}}

\author{Nazario Tantalo\\
        \\
        INFN sezione "Tor Vergata", Via della Ricerca Scientifica 1, I-00133 Rome, Italy
        \&\\
        Centro E. Fermi, Compendio Viminale, I-00184 Rome, Italy
        \\      
        E-mail: \email{Nazario.Tantalo@roma2.infn.it}}

\abstract{We compute the mass of the b-quark and the $\mrm{B_s}$ meson decay
          constant in quenched lattice QCD using a combination of HQET and the 
          standard relativistic QCD Lagrangian. We start from a small volume, where 
          one can directly deal with the b-quark, and compute the evolution
          to a big volume, where the finite size effects are negligible 
          through step scaling functions which give the change of the
          observables when $L$ is changed to $2L$. In all steps we extrapolate
          to the continuum limit, separately in HQET and in QCD for masses
          below $m_\mrm{b}$. The point $m_\mrm{b}$ is then reached by an interpolation
          of the continuum results.
          With $r_0=0.5\,\fm$ and the experimental $B_\mrm{s}$ and $K$ masses 
          we find $\fBs=191(6)$~MeV and the renormalization group invariant
          mass $M_\mrm{b}=6.89(11)$~GeV, translating into
          $\mbar_\mrm{b}(\mbar_\mrm{b})=4.42(7)$~GeV
          in the $\msbar$ scheme.
\vspace{0.7cm}\\  DESY 06-174 \\ SFB/CPP-06-47}

\FullConference{XXIV International Symposium on Lattice Field Theory\\
                July 23-28 2006\\
                 Tucson Arizona, US}

\begin{document}

\section{Introduction}

The b-quark mass is a fundamental parameter of QCD,
and its accurate knowledge is needed for theoretical predictions
of B meson decay rates. The understanding of the latter is a very active field
of high energy physics research.
At the same time the B meson decay constant plays a crucial role in the description
of these phenomena.

We focus our attention on the pseudoscalar $\mrm{B_s}$ meson, a system characterized
by two different scales: the heavy quark mass 
($m_{\rm b}\sim 5$~GeV) and the typical QCD scale. The mass of the 
strange quark is around or below the latter. We 
fix it to its physical value through $\mk$ as in \cite{mbar:pap3}.

\section{HQET and step scaling method (SSM)}

We deal with these two scales in (quenched) lattice QCD with the SSM
introduced in \cite{deDivitiis:2003iy,deDivitiis:2003wy}, but constraining the large mass behaviour
by HQET~\cite{hqet:pap1}.
The computation of an observable $O(m_{\rm h})$ 
using the SSM is based on the identity 
\be\label{eq:SSM_identity}
O(m_{\rm h},L_\infty)=O(m_{\rm h},L_0)\,
\frac{\displaystyle O(m_{\rm h},L_1)}{\displaystyle O(m_{\rm h},L_0)}\,
\cdots\, \frac{\displaystyle O(m_{\rm h},L_N)}{\displaystyle O(m_{\rm h},L_{N-1})}\,
\frac{\displaystyle O(m_{\rm h},L_\infty)}{\displaystyle O(m_{\rm h},L_N)}\,,
\ee
where $m_{\rm h}$ stands generically for a heavy quark mass whose precise
definition is needed only later.
In order to be able to extract each factor in the continuum limit, 
the starting volume $L_0$ has to be small enough to 
properly account for the dynamics of the b-quark, 
using a relativistic $\rmO(a)$-improved action.
A  good choice is $L_0=0.4$ fm ~\cite{deDivitiis:2003iy,deDivitiis:2003wy},
where easily lattice spacings of $a\approx0.012$~fm can be used.
(Physical units are set using $r_0=0.5$ fm 
~\cite{pot:r0,Necco:2001gh,Guagnelli:2002ia}).
Furthermore, $L_\infty$ has to be large 
enough such that finite size effects in $O(m_{\rm h},L_\infty)$ are negligible. 
In practise we 
will use $L_\infty \approx L_2=1.6$~fm. We will 
choose a fixed ratio $s=L_i/L_{i-1}$ in the 
step scaling functions
\be\label{eq:SSM_def}
\sigma_{O}(m_{\rm h},L_i)=
\frac{\displaystyle O(m_{\rm h},L_i)}{\displaystyle O(m_{\rm h},L_{i-1})}\,.
\ee   
The number $N$ and the scale ratio $s$ of the steps 
are in principle dependent on the considered observable 
and on the desired level of accuracy. 
It has been seen ~\cite{deDivitiis:2003iy,deDivitiis:2003wy} 
that $(N,s)=(2,2)$ is a suitable choice 
for the mass and decay constant of the ${\rm B_s}$ meson.

In HQET the step scaling functions are expanded as
\be\label{eq:SSM_expansion}
\sigma_{O}(m_{\rm h},L_i)=\sigma_{O}^{(0)}(L_i)
+\frac{\displaystyle \sigma_{O}^{(1)}(L_i)}{\displaystyle L_im_{\rm h}}
+{\rm O}\left(\frac{\displaystyle 1}{\displaystyle \left(L_im_{\rm h}\right)^2}\right)\, 
\ee  
at fixed $L_i$.
We will see that the correction terms to the 
leading order are small for the masses of interest.\\
We first consider the case of a finite volume pseudoscalar meson 
mass,  $O(m_{\rm h},L)=M_{\rm PS}(m_{\rm h},L)$, which will be 
defined in the following section.
In this case, $\sigma_{O}^{(0)}=1$ and the first non-trivial
term $\sigma_{O}^{(1)}$ 
is computable in the static approximation of HQET.
We further define
\be\label{eq:def_x}
x(m_{\rm h},L)\equiv\frac{\displaystyle 1}{\displaystyle LM_{\rm PS}(m_{\rm h},L)}
= \frac{\displaystyle 1}{\displaystyle Lm_{\rm h}} 
+ \rmO\left(\frac{\displaystyle 1}{\displaystyle \left(Lm_{\rm h}\right)^2}\right)\,,
\ee
as the natural non-perturbative dimensionless mass variable.  
The step scaling
function for the meson mass is then written as
\be\label{eq:sigma_m_2}
\sigma_{\rm m}(x,L_i)\equiv
\frac{\displaystyle M_{\rm PS}(m_{\rm h},L_i)}{\displaystyle M_{\rm PS}(m_{\rm h},L_{i-1})}=
1+\sigma_{\rm m}^{\rm stat}(L_i)\cdot x+\Or(x^2)\,, \quad
x=x(m_{\rm h},L_i)\,.
\ee
It is defined for all $x,L$. The idea for its
numerical evaluation is to compute $\sigma_{\rm m}^{\rm stat}(L)$ explicitly
in the static approximation and fix the small remainder 
by the relativistic QCD data with quarks of masses
of the physical charm quark and higher.
In other words we interpolate to the physical b-quark mas. With the experimental mass
of the $\mrm{B_s}$ meson, $\MBs=5.3675(18)\,\GeV$ we fix 
$
  x_2 = 1/L_2\MBs 
$
and the physical points corresponding to the b-quark are then
given by
\be\label{eq:x1_star}
  x_2 = 1/(L_2\MBs)\,,\quad x_{i-1}=2\sigma_{\rm m}(x_i,L_i)\cdot x_i\,.
\ee
The numerical results will have to be
evaluated at these points.
In the smallest volume we relate the meson mass to the renormalization group
invariant (RGI) quark mass, $M_{\rm h}$, defining 
\be\label{eq:rho}
\rho(x,L_0) \equiv
\frac{\displaystyle M_{\rm PS}(m_{\rm h},L_0)}{\displaystyle M_{\rm h}}=
\rho^{(0)}(L_0)+\rho^{(1)}(L_0)\cdot x+\Or(x^2)\,.
\ee
We thus have the connection of the $\mrm{B_s}$ meson mass and the RGI b-quark mass
\be\label{eq:b_mass}
M_{\rm b}=\frac{\displaystyle \MBs}{\displaystyle \rho(x,L_0)\cdot 
          \sigma_{\rm m}(x_1,L_1)\cdot \sigma_{\rm m}(x_2,L_2)}\,.
\ee 
For the decay constant the step scaling function 
\bea
\sigma_{\rm f}(x,L_i)&\equiv&
\frac{\displaystyle f_{\rm PS}(m_{\rm h},L_i    )\sqrt{M_{\rm PS}(m_{\rm h},L_i)}}{
      \displaystyle f_{\rm PS}(m_{\rm h},L_{i-1})\sqrt{M_{\rm PS}(m_{\rm h},L_{i-1})}}
 =\sigma_{\rm f}^{\rm stat}(L_i)
  +\sigma_{\rm f}^{(1)}(L_i)\cdot x+\Or(x^2) \label{eq:sigma_dc}\,
\eea
yields straightforwardly the connection between the finite volume
decay constant and the infinite volume one. Note that the only
approximation made in the above equations is to neglect 
finite size effects on mass and decay constant in the volume 
of linear extent $L_2$.

\section{Finite volume observables}

\subsection{Relativistic QCD}

Suitable finite volume observables are defined in the QCD  \SF ~\cite{SF:LNWW,SF:stefan1}
with a space-time topology $L^3\times T$,
where $T=2L$ and $C=C'=0$ is chosen for the boundary gauge fields, and $\theta=0$ for 
the phase in the spatial quark boundary conditions.\\
The \Oa-improved correlation functions $f_{\rm A}(m_{\rm h},L,x_0), f_{\rm P}(m_{\rm h},L,x_0)$ 
and $f_1(m_{\rm h},L)$ are defined and renormalized as in ~\cite{deDivitiis:2003iy},
allowing to compute the pseudoscalar meson decay constant
\be\label{eq:dc_qcd}
f_{\rm PS}(m_{\rm h},L)=\frac{\displaystyle -2}{\displaystyle \sqrt{L^3 M_{\rm PS}(m_{\rm h},L)}}
\frac{\displaystyle f_{\rm A}(m_{\rm h},L,L)}{\displaystyle \sqrt{f_1(m_{\rm h},L)}}\stackrel{m_{\rm h}\to m_{\rm b}}{=}
\fBs(L)\stackrel{L\to\infty}{=}\fBs
\ee 
and the pseudoscalar meson mass
\be\label{eq:meson_mass_qcd}
M_{\rm PS}(m_{\rm h},L)=\frac{\displaystyle 1}{\displaystyle 2a}
\ln{\left[\frac{\displaystyle f_{\rm A}(m_{\rm h},L,L-a)}{\displaystyle f_{\rm A}(m_{\rm h},L,L+a)}\right]}
\stackrel{m_{\rm h}\to m_{\rm b}}{=}\MBs(L)\stackrel{L\to\infty}{=}\MBs
\ee
For all observables computed in relativistic (quenched) QCD
we employ the non-perturbatively \Oa-improved Wilson action\cite{impr:pap1,impr:pap3}. 
The data at finite heavy quark mass were published
in ~\cite{deDivitiis:2003iy,deDivitiis:2003wy}.  
They have been reanalyzed, taking into 
account the correlation between observables computed on the 
same gauge configurations. 
The statistical uncertainties on the 
renormalization constants and the lattice spacing are included before 
performing the continuum limit extrapolations; they do not appear as a 
separate uncertainty.

\subsection{HQET}

In the static approximation of HQET,
unrenormalized correlation functions $\fastat$ and $\fonestat$ 
are defined in complete analogy to the relativistic ones,
see \cite{Heitger:2003xg}. As in this reference,
we use the RGI
static axial current, related to the bare one
by a factor $\ZRGI$. It serves to define the RGI ratio\,,
\be
\YRGI(L)=\ZRGI\frac{\displaystyle \fastat(L,L)}{\displaystyle\sqrt{\fonestat(L)}}\,,
\ee
which is related to the QCD decay constant $f_{\rm PS}$ via
\be\label{eq:fb_hqet}
f_{\rm PS}(m_{\rm h},L)\sqrt{L^{3}M_{\rm PS}(L)}=
 -2\Cps(\Lambda_\msbar/M_{\rm h})\times\YRGI(L)+\rmO(1/m_{\rm h})\,.
\ee
The function $\Cps(\Lambda_\msbar/M_{\rm h})$, defined in \cite{Heitger:2003xg},
can be accurately evaluated  in perturbation theory; we use the 3-loop
anomalous dimension $\gamma^{\rm PS}$ computed in \cite{Chetyrkin:2003vi}.
Just like $\ZRGI$,
it is needed only for $f_{\rm PS}(m_{\rm h},L_0)$; 
it cancels out in the step scaling functions.

In analogy to \eq{eq:meson_mass_qcd} we further define
$
\Gamma_{\rm stat}(L)=\frac{1}{ 2a}
\ln\left[{ \fastat(L,L-a)}/{\fastat(L,L+a)}\right]\,.
$
The static step scaling functions then read
\be\label{eq:stat_sig}
\sigma_{\rm f}^{\rm stat}(L_i)=\frac{\displaystyle 1}{\displaystyle 2^{3/2}}
\frac{\displaystyle \YRGI(L_i)}{\displaystyle \YRGI(L_{i-1})},\quad
\sigma_{\rm m}^{\rm stat}(L_i)=L_i\,[\Gamma_{\rm stat}(L_i)-\Gamma_{\rm stat}(L_{i-1})]\,,
\quad L_i = 2L_{i-1}\,.
\ee
These quantities will be precisely computed 
by using the static action denoted by HYP2 in \cite{DellaMorte:2005yc} 
(see also \cite{HYP}),
and the corresponding $\Oa$-improvement coefficients for the static axial current.
The regularization independent part of the factor $\ZRGI$ is known 
from \cite{Heitger:2003xg}, while the regularization dependent one 
is computed in this work.

\section{Numerical results for the $\mathbf{b}$ quark mass}

\begin{figure}[t]
\vspace{-0.4cm}
\begin{center}
\begin{tabular}{cc}
\includegraphics[scale=0.4]{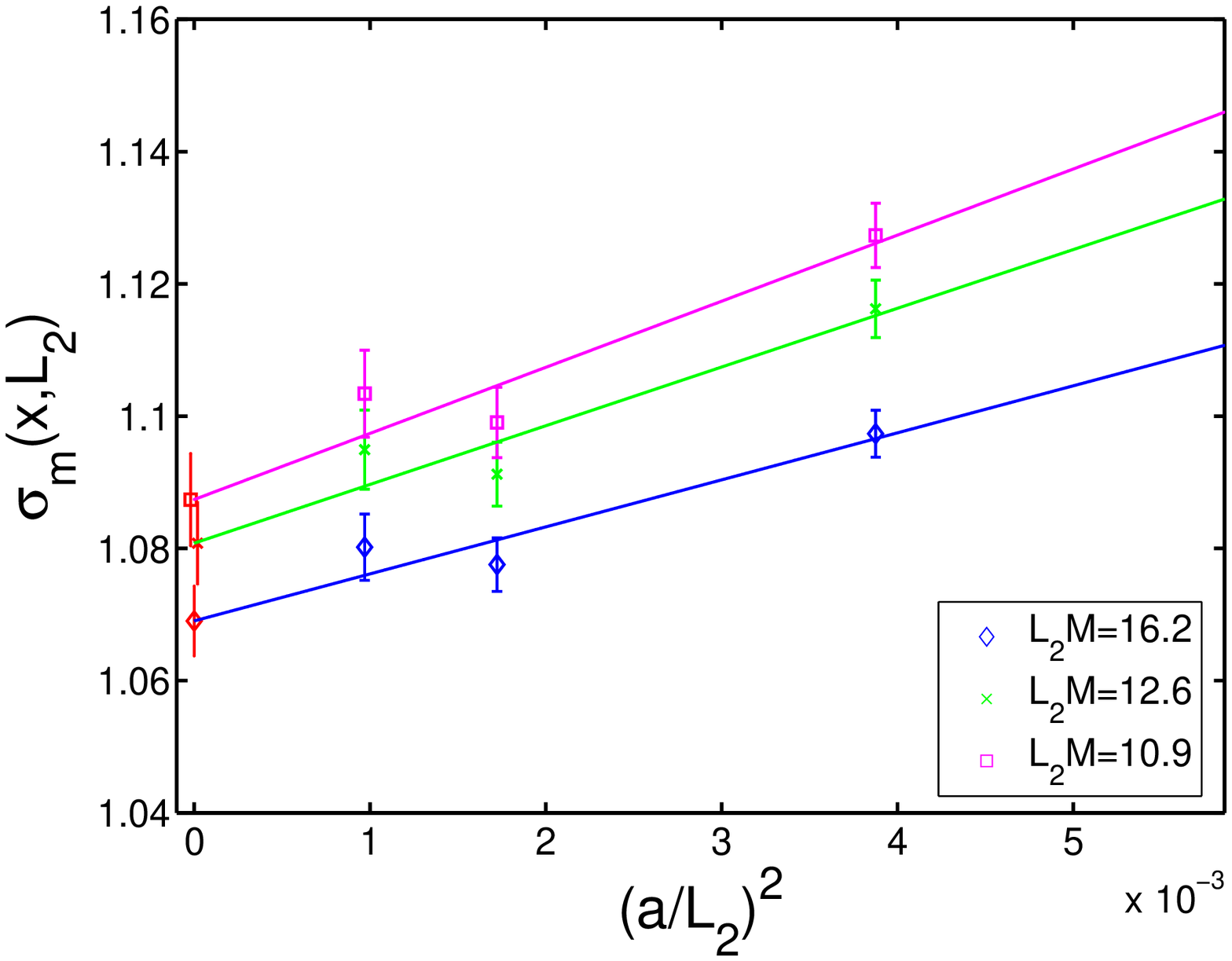} &
\includegraphics[scale=0.4]{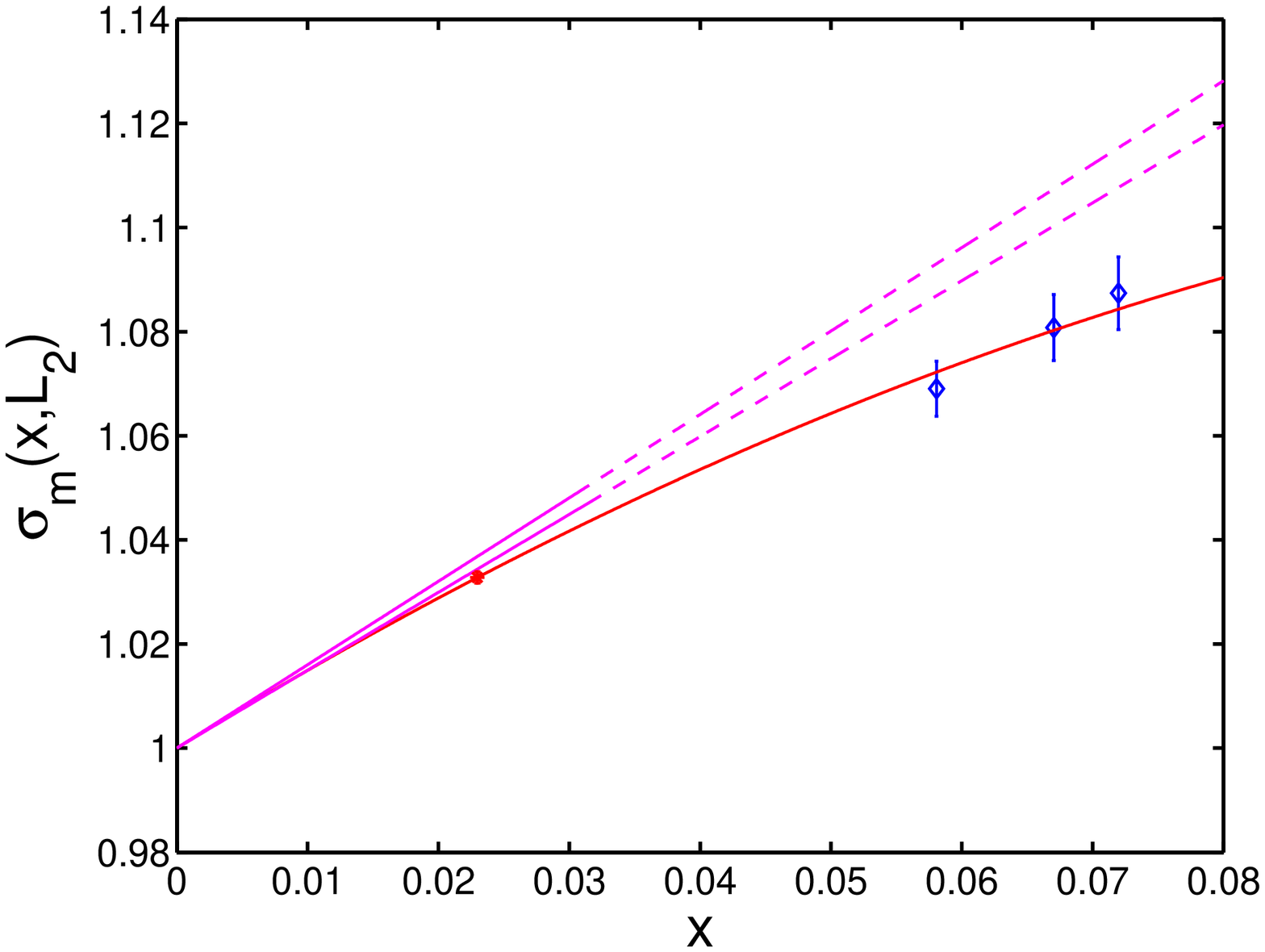} \qquad \\
\end{tabular}
\end{center}
\vspace{-0.5cm}
\caption[]{\label{fig:sigma2_m}
{Continuum limit extrapolation and interpolation 
of $\sigma_\mrm{m}(x,L_2)$}}
\end{figure}

The computation of  $\sigma_{\rm m}(x,L_2)$ is performed
at finite quark mass on lattices with $\beta=5.960$,
$6.211,\ 6.420\ $ and 
resolutions $L_2/a=16,\ 24,\ 32$; the continuum limits for the three 
heaviest quark masses are shown on the left of \Fig{fig:sigma2_m}.
For the static step scaling function we took the results for $L=L_2$ 
from an extension\cite{Estat:me} of the work of the ALPHA collaboration\cite{stat:letter},
while in the intermediate volume ($L_1$) we simulated lattices with $5.960\leq\beta\leq6.737$. 
The continuum limit 
\be\label{eq:CL_S2_stat}
\sigma_{\rm m}^{\rm stat}(L_2)=1.549(33)\,,
\ee
is used in the interpolation of $\sigma_{\rm m}(x,L_2)$ 
between values of $x$ corresponding to about the mass
of the charm quark and the limit $\sigma_{\rm m}(0,L_2)=1$.
It constrains the 
slope of the fitting curve to the cone
shown in  \Fig{fig:sigma2_m}. The result of the quadratic fit
in $x$  reads
\be\label{eq:res_s2m}
\sigma_{\rm m}(x_2,L_2)=1.0328(11)\,,
\ee
hardly distinguishable from a purely static result.
Analogously the interpolation of the step scaling function 
for the intermediate volume gives 
\be\label{eq:res_s1m}
\sigma_{\rm m}(x_1,L_1)=1.0092(18)\,.
\ee

In the small volume only the relativistic data are needed to establish 
a finite volume relationship between the meson and the heavy quark masses. 
The renormalization is non-perturbatively achieved through the 
renormalization factor $\zM(g_0)$ and the 
$\Oa$-improvement terms computed in  \cite{Capitani:1998mq,impr:babp,Heitger:2003ue}. 
Using \eq{eq:b_mass}, the interpolated value
\be\label{eq:res_rho}
\rho(x_0,L_0)=0.748(11)\,,
\ee
is combined with the above step scaling functions to 
find the scale and scheme independent number
\be\label{eq:res_b_mass}
M_{\rm b}=6.888(105)\,\GeV\hspace{0.3cm}\Rightarrow\hspace{0.3cm}\mbMSbar(\mbMSbar)=4.421(67)\,\GeV.
\ee

\section{Numerical results for the decay constant}

\begin{figure}[t]
\vspace{-0.4cm}
\begin{center}
\begin{tabular}{cc}
\includegraphics[scale=0.4]{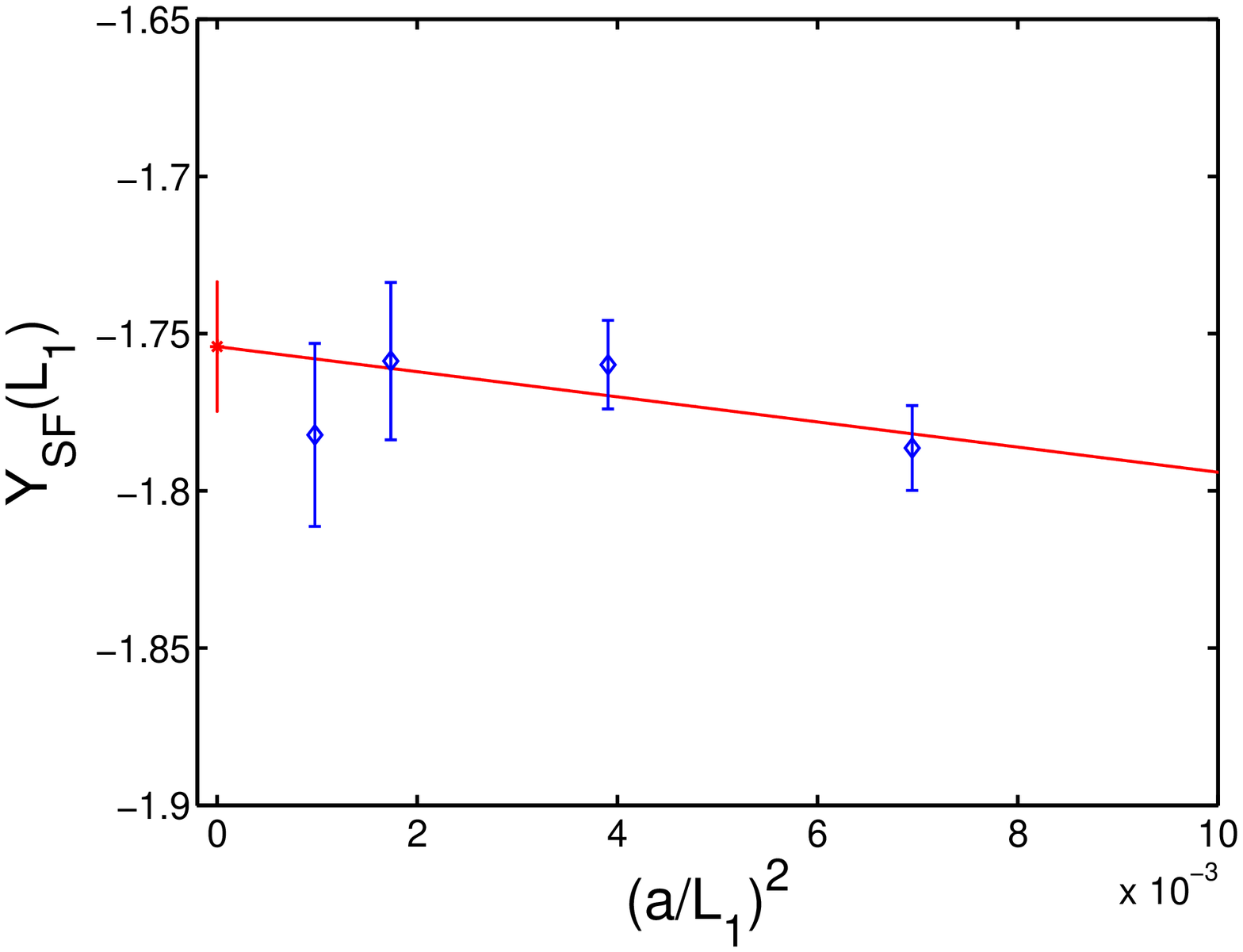} &
\includegraphics[scale=0.4]{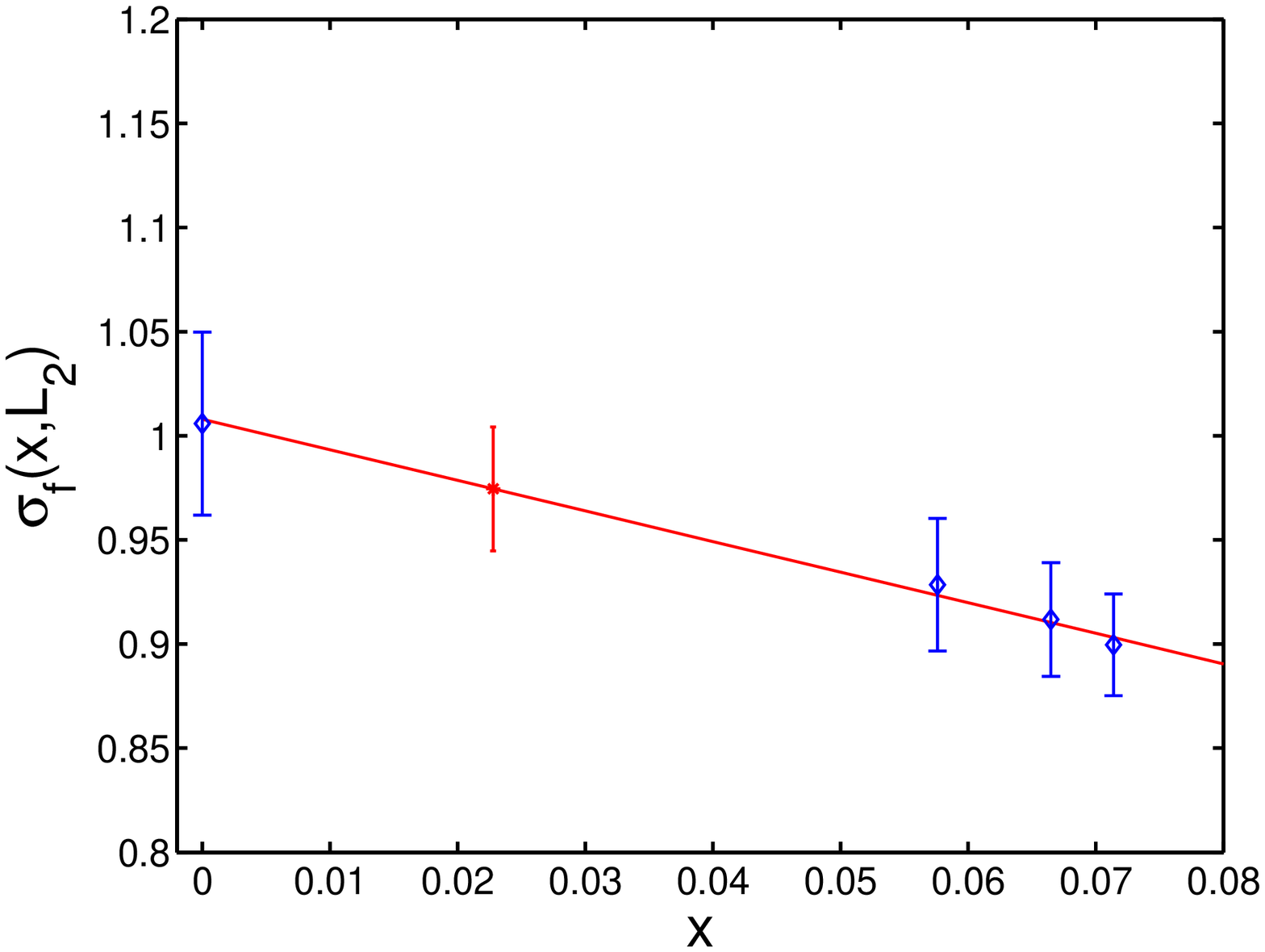} \qquad \\
\end{tabular}
\end{center}
\vspace{-0.5cm}
\caption[]{\label{fig:sigma2_dc}
{Continuum extrapolation of $Y_{\rm SF}(L_1)$ and interpolation 
of $\sigma_\mrm{f}(x,L_2)$}}
\end{figure}

For the computation of $\sigma_{\rm f}(x,L_2)$ the relativistic data originate from 
the same gauge configurations used earlier, while in the static case
the decay constant in the bigger volume, 
\be\label{eq:dc_stat_Linfty}
\YRGI(L_2)=-4.63(19)\,,
\ee
was again computed and extrapolated to the continuum limit as an
extension \cite{Estat:me} of \cite{stat:letter}.
The continuum extrapolation of the
same quantity in the intermediate volume ($L=L_1$)
is shown on the left of \Fig{fig:sigma2_dc}. The result
\be\label{eq:YRGI_2L0}
\YRGI(L_1)=-1.628(19)
\ee
is used together with (\ref{eq:dc_stat_Linfty}) and the relativistic data, 
as shown on \Fig{fig:sigma2_dc} (right), to get
\be\label{eq:S2}
\sigma_{\rm f}^{\rm stat}(L_2)=1.006(44),\hspace{1.0cm}
\sigma_{\rm f}(x_2,L_2)=0.974(30)\,.
\ee
Similarly, but by extrapolating the step scaling function to the
continuum limit rather than $\YRGI(L_1)$ and $\YRGI(L_0)$ separately,
we obtain
\be\label{eq:S1}
\sigma_{\rm f}^{\rm stat}(L_1)=0.4337(44),\hspace{1.0cm}
\sigma_{\rm f}(x_1,L_1)=0.4260(31)\,.
\ee
{\flushleft With the small volume results (see \Fig{fig:sv_dc})}
\be\label{eq:SV_DC}
\YRGI(L_0)=-1.347(13),\hspace{1.0cm}
Y_{\rm PS}(x_0,L_0)=
\frac{\displaystyle -\fBs(L_0)\sqrt{L_0^3 M_{\rm PS}(L_0)}}{\displaystyle 2\Cps(\Lambda_\msbar/M_{\rm b})}
 =-1.280(17)\,,
\ee
we finally arrive at the result
\be\label{eq:res_fbs}
\fBs=191(6)\,\MeV \,.
\ee
\section{Conclusions}

\begin{figure}[t]
\vspace{-0.4cm}
\begin{center}
\begin{tabular}{cc}
\includegraphics[scale=0.4]{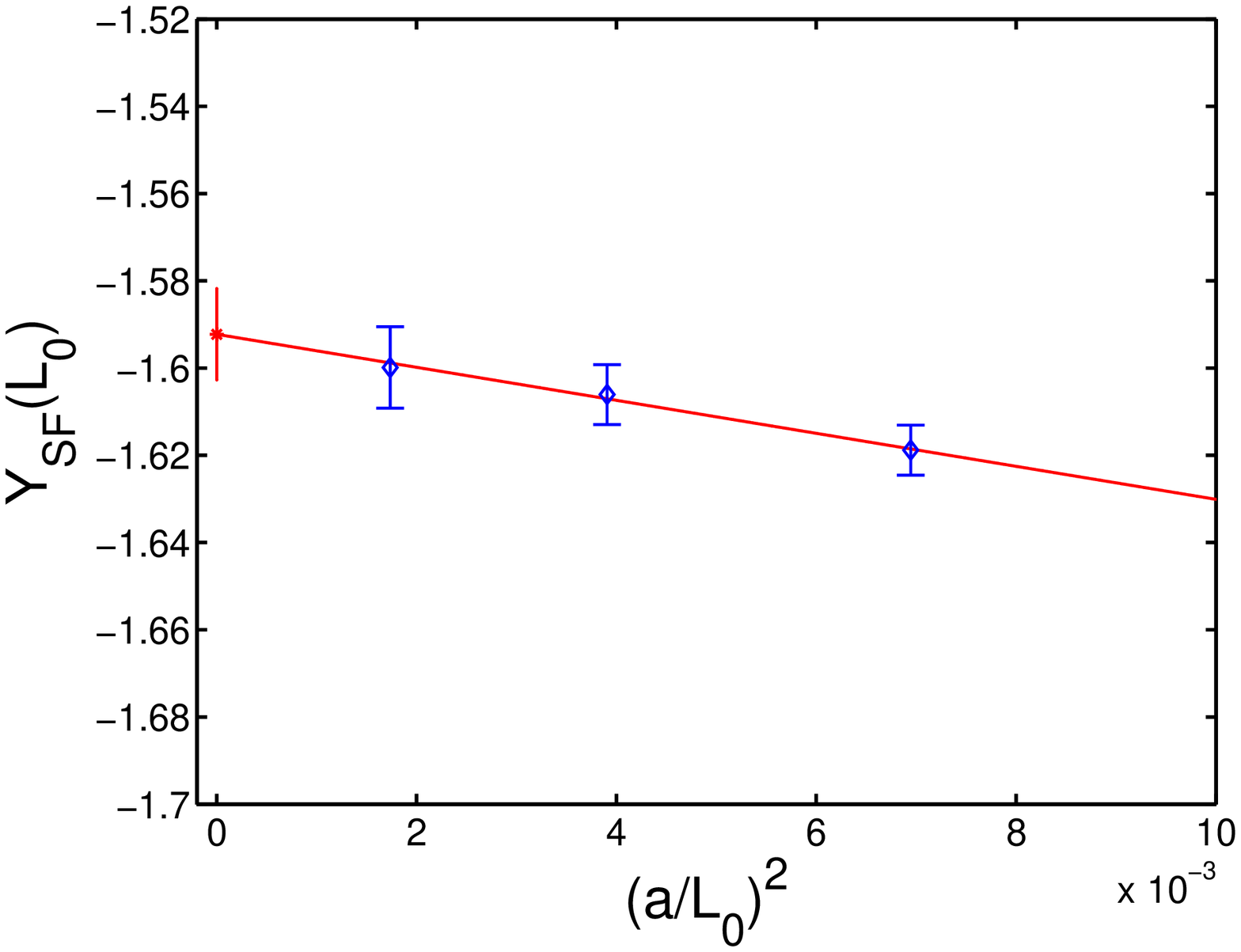} &
\includegraphics[scale=0.4]{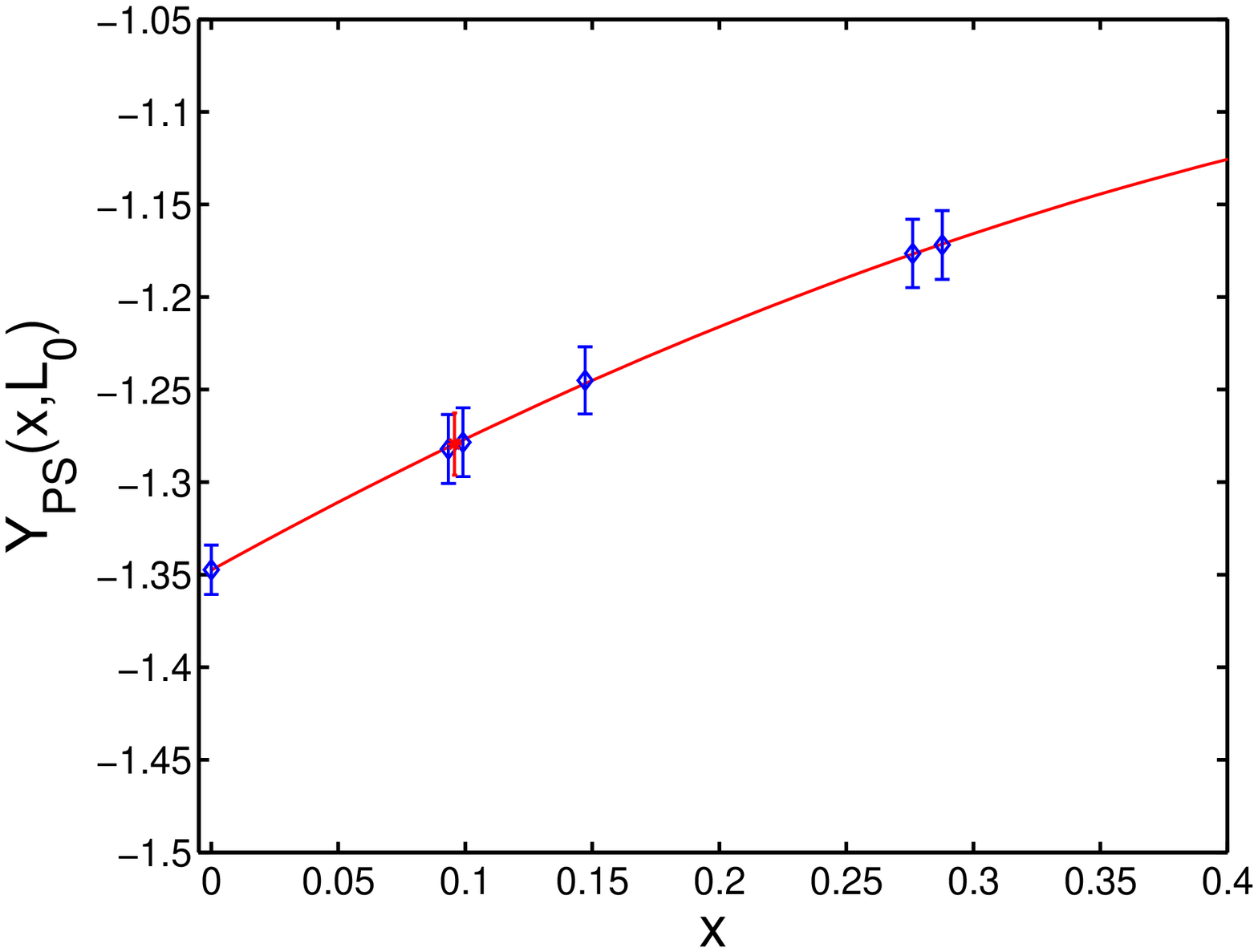} \qquad \\
\end{tabular}
\end{center}
\vspace{-0.5cm}
\caption[]{\label{fig:sv_dc}
{Continuum extrapolation of $Y_{\rm SF}(L_0)$ and interpolation 
of the decay constant on the small volume}}
\end{figure}

The combination of the Tor Vergata strategy to 
compute properties of heavy-light mesons ~\cite{deDivitiis:2003iy,deDivitiis:2003wy}
with the expansion of all quantities in HQET~\cite{hqet:pap1}, changes extrapolations
in the former computations into interpolations. As expected, our numerical
results demonstrate that these are very well behaved.
Indeed the higher order mass dependence of the step scaling functions is very weak,
and in all but one steps the static approximation alone gives very accurate
results. In the one exception ($Y_{\rm PS}(x_0,L_0)$, \fig{fig:sv_dc}) the $\rmO(1/\mbeauty)$ 
corrections are around 5\%. 

Our results do not suffer from any systematic errors
apart from the use of the quenched approximation; small systematic errors quoted
in ~\cite{deDivitiis:2003iy,deDivitiis:2003wy} for the extrapolation uncertainties 
have been eliminated. Our results are in agreement with the ones of
~\cite{deDivitiis:2003iy,deDivitiis:2003wy,stat:letter,mb:nf0},
within the errors.\\
Concerning dynamical fermion computations, the challenge in this strategy
is to simulate in a large volume (such as $L_2$) with small enough lattice spacings, 
where quark masses of
around $\m_{\rm charm}$ and higher can be simulated with confidence.\\

\noindent {\bf Acknowledgement.}
We thank Michele Della Morte, Stephan D\"urr, Jochen Heitger and Andreas J\"uttner
for useful discussions and the permission to use results of \cite{Estat:me} prior to publication.


\bibliographystyle{h-elsevier}   
\bibliography{refs}           

\end{document}